\documentstyle[epsfig,twocolumn,subfigure]{mn2e}
\newcommand{\goodgap}{\hspace{\subfigtopskip} \hspace{\subfigbottomskip}}

\title[Rotation Curves and nonextensive statistics]{Rotation Curves and Nonextensive Statistics}
\author[C. Frigerio Martins, J.A.S. Lima, and P. Chimenti]
{C. Frigerio Martins$^{1}$, J.A.S. Lima$^{1}$, P. Chimenti$^{2}$\\
$^{1}$IAG/USP, Instituto de Astronomia, Geof\'{i}sica e Ci\^{e}ncias Atmosf\'{e}ricas/Universidade de S\~{a}o Paulo\\
Rua do Mat\~{a}o 1226. 05508-090 S\~{a}o Paulo, Brazil\\
$^{2}$UFABC, Universidade Federal do ABC\\
Av. dos Estados 5001. 09210-580 Santo Andr\'{e}, Brazil\\
uelchris@hotmail.com, jas.lima@iag.usp.br, pietro.chimenti@ufabc.edu.br
}

\date{Accepted xxx, Received yyy, in original form zzz}

\begin{document}

\maketitle

\begin{abstract}

We investigate the influence  of the nonextensive $q$-statistics and kinetic theory on galactic scales through the analysis of a devised sample of Spiral Rotation Curves.
Largely supported by recent developments on the foundations of statistical mechanics and a plethora of astrophysical applications,  the theory also 
provides an alternative interpretation to the empirical cored dark matter profiles observed in galaxies.
We show that the observations could well be fitted with reasonable values for the mass model parameters, encouraging further investigation into $q$-statistcs on the distribution of dark matter from both observational and theoretical points of view.

\end{abstract}


\section{Introduction}

The nature of dark matter (DM) is a challenging unsolved problem in astronomy and even
its density distribution on the galactic scales is still under debate.
In the last few decades was also established that the kinematics of disk galaxies exhibit a mass discrepancy \cite{bosma79,rubin80}.
Spirals show an inner baryon dominance region
\cite{athanassoula87,persic88,palunas00},
whose size ranges between 1 and 3 disk exponential length-scales according to the galaxy luminosity 
\cite{salucci99},
inside which the observed ordinary baryonic matter accounts for the rotation curve (RC).
However, outside the characteristic disk region, the distribution of the baryonic components cannot justify the observed profiles and the amplitudes of the measured circular velocities
\cite{bosma81,gentile07a,gentile07b}.
This puzzle is usually solved by adding an extra mass component, the so-called DM halo.
RCs have primarily been used to assess the existence, the amount and the distribution of this dark component.

The study of the systematics of spiral kinematics 
\cite{persic91,persic96,salucci07}
has evidenced that these systems present universal features well correlating with global galactic properties, leading to the construction of an empirical function of galactocentric radius, dubbed the Universal Rotation Curve (URC), that, tuned by a global galaxy property (e.g. luminosity), is able to reproduce the RC of any object, out to the virial radii.
It is the observational counterpart of the velocity profile that emerges out of CDM large N-body numerical simulations of structure formation.
As individual RCs, it implies a mass model including a Freeman disk and a DM halo with an empirical Burkert cored profile
\cite{burkert95,salucci00}

\begin{equation}\label{eq1}
\rho_{B}(r)=\frac {\rho_0 r^3_0}{(r+r_0)(r^2+r^2_0)},
\end{equation}
where $r_0$ is the core radius and $\rho_0$  its central density.

Recent debate on the literature has focused on the cuspiness of the DM  density distribution in the centers of galaxy halos as predicted from the simulations
\cite{nfw96,moore99,navarro04,neto07},
commonly represented by the Navarro, Frenk and White profile, but not seen in observed data, as well as in the various systematics of the matter distribution \cite{deBlok01,deBlok02,marchesini02,gentile04,gentile05,gentile07a,salucci07,deBlok10,naray11}.
 
In principle, the rotation curves should be fitted based on a more fundamental approach, as for instance, by changing the underlying kinetic description of DM halos. 
In this concern, we recall that a cored density profile of DM halos like Eq. 1  cannot be explained by a maximum finite phase space density
\cite{burkert95}.
An alternative possibility to be investigated here is provided by the $q$-statistics proposed long ago by Tsallis (1988), where the $q$-parameter  quantifies the degree of nonextensivity. This nonadditive extension of the standard Boltzmann-Gibbs (BG) approach is remarkably successful 
and widely applied in a  variety of different phenomena in condensed matter, plasma physics and, more recently, in astronomy
\cite{tsallis14}.
Indeed, even a rigorous nonextensive kinetic approach has been developed
\cite{silva98,lima00,lima01}
and applied to high energy physics \cite{wong13} and astrophysical problems
\cite{taruya02,lima05,hansen05,burlaga06,betzler12}.

In this work we explore the roots of the cored DM profiles based on the  predictions of the $q$-kinetic theory for a collisionless self-gravitating particle distribution  on galactic scales
\cite{lima05}
thereby  investigating the consistency of the resulting halos concerning the fitting of the RCs of spirals.
There are some few attempts in literature for describing the DM halos in this context, as for instance \cite{leubner05,kronenberger06},
and more recently a mass modeling of one galaxy \cite{cardone11}.
However, as far as we know, this is the first time that a complete test with a proper wide sample is performed. As we shall see, the nonextensive distribution provides an appealing way to describe DM cored halos from first principles.

The paper is organized as follows.
In Section 2, the  main theoretical results of the $q$-kinetic theory which are relevant for our data analysis are briefly discussed.
In Section 3, we present our sample and methodology of analysis and in Section 4 the main results of the article are obtained.
Finally, Section 5 summarizes the basic findings.

\section{Why nonextensive statistics and kinetic theory?}

It is widely known that
the long-range nature of the gravitational force is incompatible with the underlying assumptions
of BG statistics and the associated kinetic theory, where the interactions among particles (binary collisions) are assumed to
be short-ranged with the entropy obeying the so-called additivity principle. In order to cure this kind of problem,  Tsallis (1988) proposed a nonextensive extension  of the BG approach based on the following entropic functional for a discrete system  with W accessible microstates:

\begin{equation}\label{eq2} 
S_q = k_B \frac {\sum_{i}^{W}
p_i^{q} - 1}{1-q},
\end{equation} where $k_B$ is the
Boltzmann constant, $q$ is the entropic-index quantifying the degree of
nonextensivity,  and $p_i$ is the probability of the $i$th
microstate. This expression  is a one parametric formula adopted 
to extend the applicability of BG statistical
mechanics in the presence of long range interactions. It reduces to the 
standard BG entropy  in the limiting case ($q = 1$) since for equal probabilities, $p_i =1/W$, one finds
\begin{equation}\label{eq3}
\lim_{q \to 1} S_{q} = \lim_{q \to 1} k_B \frac {\sum_{i}^{W}
p_i^{q} - 1}{1-q} = k_B\ln W .
\end{equation}

On the other hand, given a composite system $A+B$ which are independent in the sense that $p_{ij}^{A+B} = p_i^A p_j^B$,  
one may show that the total entropy satisfies, $S_q(A+B)=S_q(A)+S_q(B)+ k_B^{-1}(1-q)S_q(A)S_q(B)$. Hence, 
if $q$ differs from unity, the entropy becomes nonextensive. It is 
superextensive if $q < 1$, subextensive when $q > 1$ with the
Boltzmann factor generalized into a power law.

Such results are readily extended to the continous case with the finite sum replaced by an integral. 
For a collision (or collisionless) gaseous system, for instance, the main implication of the above extension is that the canonical exponential Maxwell-Boltzmann factor is replaced by a power law (see, for instance, Silva, Plastino \& Lima 1998; Lima, Plastino \& Silva 2001). 

Nonextensive kinetic theory has been applied to many problems in plasma physics, as well in the astrophysical context. For instance, dispersion relations for electrostatic plane-wave propagation in a collisionless thermal plasma were discussed in the last decade (Lima, Silva \& Santos 2000; Silva, Alcaniz \& Lima 2004). A similar technique was also applied to the reanalyse the celebrated gravitational Jeans instability mechanism. It was found that for power-law distributions with cutoff,
the instability condition is weakened with the system becoming unstable even for wavelengths of the disturbance smaller
than the standard Jeans length \cite{lima02}. Interesting studies are related to the solar neutrino problem,  fluxes of cosmic rays and gravothermal catastrophe in stellar systems, among others \cite{duJiulin04,abelev13,tsallis14}.

Analytical and numerical investigations of the velocity distribution function for self-gravitating collisionless particles including dark matter and star clusters were discussed by some authors
\cite{hansen05,lima05,cardone11}.
In particular, Lima \& de Souza 2005 computed the density profiles and other quantities of physical interest assuming that spherically symmetric collisionless systems may relax to the non-Gaussian power-law distribution as suggested by the nonextensive kinetic theory.
In their analysis, the density profile  reads:

\begin{equation}\label{eq4}
\rho = \rho_q [1-(1-q)\theta]^{\frac{5-3q}{2(1-q)}},
\end{equation}
where $\rho_q$ is a constant density which depends on the $q$-index while the  $\theta$ function,
which is defined in terms of the gravitational potential and the dispersion velocity $\theta = -\sigma^{-2}\phi(r)$, satisfies a Lane-Emden type differential equation:

\begin{equation}\label{eq5}
\frac{1}{x^2}\frac{d}{dx}(x^2 \frac{d\theta}{dx})= -[1 +
(1-q){\theta}]^{\frac{5-3q}{2(1-q)}},
\end{equation}
with  $x = r/r_o$ being a suitable dimensionless 
quantity ($r_o ={\sqrt 4\pi G \rho/\sigma^{2}}$).

It should be stressed that the  resulting  spherically symmetrical density profile obtained from the above equations may describe  a star like gaseous system, as the one appearing in globular clusters (or elliptical galaxies), as well as the dark matter halos of spiral galaxies.
Hence, in the next sections it will be assumed that the halos of dark matter obey the above coupled system. We emphasize that such a description has not been adopted neither to the stellar disk nor the gas component.

\section{Data Analysis and Methodology}

In what follows we consider the sample of selected galaxies already discussed by Frigerio Martins \& Salucci (2007).  
It includes nearby galaxies of different surface brightness representing suitable available RCs to study the mass distribution of luminous and dark matter and it has been previously used in works concerning modifications of gravity and the  core/cusp controversy 
(such galaxies present a very small bulge so that it can be neglected in the mass model to a good approximation).
The associated RCs are smooth, symmetric and extended with the intrinsic luminosity profiles well measured (for details see Frigerio Martins \& Salucci (2007) and references therein).

To begin with, let us decompose the total circular velocity into stellar, gaseous and dark matter halo contributions
\begin{equation}
V^2_{tot}(r)=r \frac{d}{dr} \phi_{tot}=V^2_{\star}+V^2_{gas}+V^2_{DM} \label{eq:Vtot}.
\end{equation}
The Poisson equation relates the surface (spatial) densities of these components to the corresponding gravitational potentials.
Available photometry shows that the stars in our sample are distributed in a thin disk, with the usual exponential surface density profile 
$\Sigma_{\star}(r)=(M_D/2\pi R^2_D) e^{-r/R_D}$
\cite{freeman}.
$R_D$ is the scale length, measured directly from the optical observations, and $M_D$ is the stellar disk mass, kept as a free parameter of our analysis.
The circular velocity contribution is given by
$V^2_{\star} (r)=(G M_D /2R_D) x^2 B(x/2)$,
where $x=r/R_D$ and $G$ is the gravitational constant.
The quantity
$B=I_0 K_0 - I_1 K_1$ 
is a combination of Bessel functions.
The contribution of the gaseous disk is directly derived from the HI surface density distribution
(few galaxies of the sample present a very small amount of gas, and for this reason have been neglected in the analysis).

In our analysis, {Eqs. (\ref{eq4}) and (\ref{eq5})} are numerically solved in order to obtain the dark matter density distribution $\rho$.
In addition, it will also be assumed a constant  dispersion velocity parameter $\sigma$  (see e.g., Cardone, Leubner \& Del Popolo 2011 and references therein).
The contribution from the DM halo to the total velocity is 
$V^2_{DM}=G M(r)/r$, where the mass profile is given by
$M(r)=4 \pi \int_0^r x^2 \rho(x) dx$. 
It has three parameters, namely, the nonextensive $q$ parameter of the theory  plus the scaling ones, which  depend on the particular self-gravitating system being studied, the core radius $r_o$ and its central density $\rho_q$.

In a first step, the RCs are $\chi^{2}$ best-fitted with the following set of free parameters:
the nonextensive term ($q$) and the scaling parameters ($r_o$ and $\rho_q$) of the theory, and the stellar disk mass ($M_D$).
The errors for the best fit values of the free parameters are calculated at one standard deviation with the $\chi^2_{red}+1$ rule.
From the results of these fits we get a mean value of $q=0.85 \pm 0.35$. 
In the second step we redo  the best-fit thereby fixing the nonextensive parameter at $q =0.85$ and keeping the quantities $r_o$, $\rho_q$ and $M_D$, as free parameters.
In our analysis the value $q=0.85$ is the most favorable for explaining the RCs of the adopted sample.
It should be noticed  that the entropic  index is limited on the interval [0,2].
The lower limit is guaranteed by the second law of thermodynamics
\cite{lima01} while the upper limit comes from studies in different fields, see
\cite{kaniadakis05,silva05}
and references therein.

\section{Results}

In Figure 1 and Table 1  we summarize the main results of our analysis.
In general, we  have determined  for all galaxies: 
(i) the velocity model $V_{tot}$ well fitting the RCs and
(ii) acceptable values for the free parameters.

The residuals of the measurements with respect to the best-fit mass model are in most of the cases compatible with the error-bars.
We also find acceptable values for the B-band mass-to-light ratio parameter for the galaxies, for which we should have approximately $0.5<\Upsilon_{\star}^{B}<6$ and a positive correlation between 
B-luminosity and \footnote{$\Upsilon_{\star}^B \equiv M_{D}/L_{B}$; $L_{B}$ is the B-band galaxy luminosity.}$\Upsilon_{\star}^{B}$
\cite{salucci08}.

In Figure 2, we display the relation between the core radii and 
central densities provided by the theoretical nonextensive DM halos in combination with our analysis of RCs.

It is interesting  that the free dynamic parameters can be reproduced by  the simple relation 
log $\rho_q$ $\simeq$ -$\alpha$ log $r_o$ + $const$,
where the slope $\alpha$ $\in$ ($0.9,1.1$) with tighter constraints requiring more accurate data.
This result is fully consistent with the ones found in the literature when the 
parameters of the empirical Burkert DM halos are obtained from dynamical modeling of spirals, weak lensing shear and the Local Group dSphs
\cite{salucci12}. Also in line with their findings, the central surface density we found, given by the quantity
$\rho_q r_o$, is nearly constant.

In Figure 3, we show a specific DM halo density distribution obtained from the galaxy ESO 116-G12 (black) in comparison with that obtained from the URC (red) with same virial mass.
We see that the model analyzed in this article provides results in almost perfect agreement  with  the cored DM halo distributions which reproduce the observed properties of spiral galaxies. Actually, there are a number of examples giving an  excellent match with the popular empirical  Burkert profile (see introduction).
As it appears, the case displayed in Fig. 3 suggests that the nonextensive approach  as discussed by Lima \& de Souza (2005) can be thought as a possible theoretical realization for such a profile.

\begin{figure*} 
\centering
\subfigure{\includegraphics[width=4.06cm]{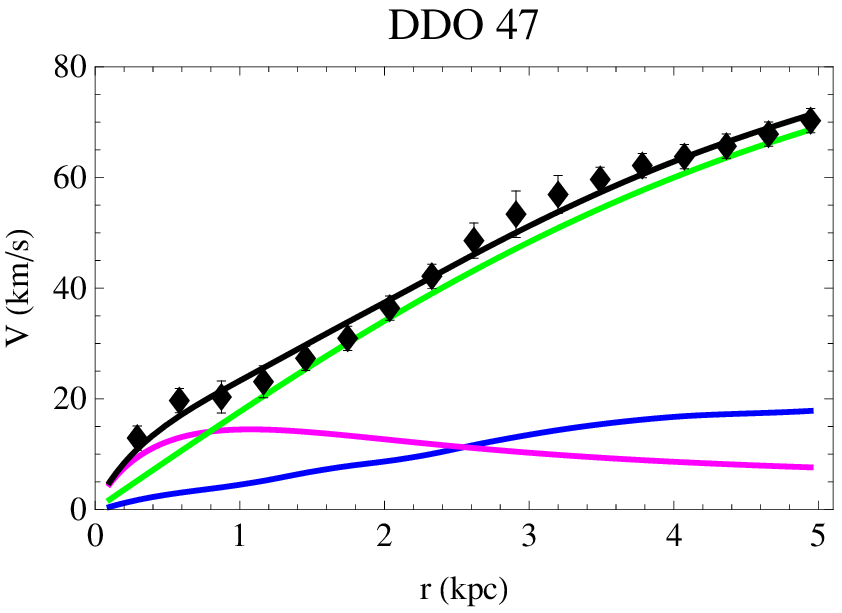}}\goodgap
\subfigure{\includegraphics[width=4.06cm]{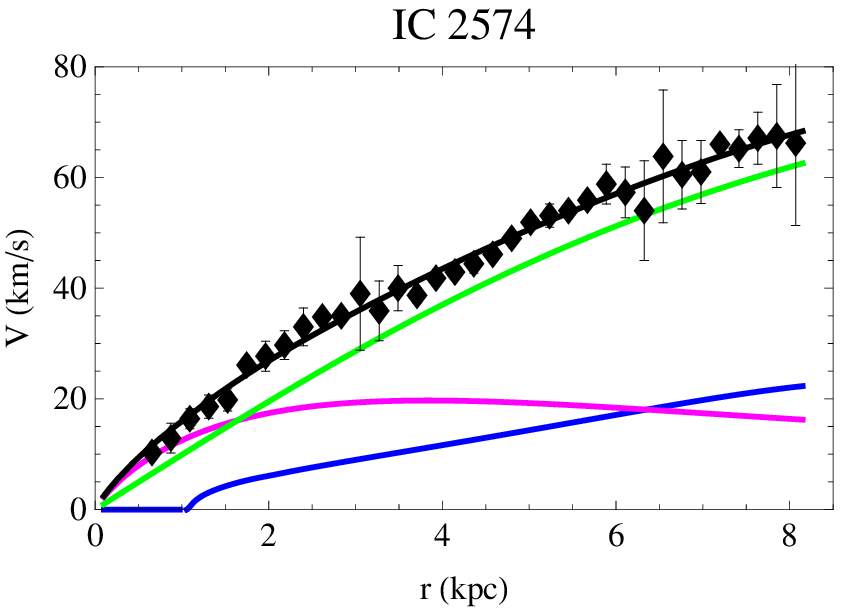}}\goodgap
\subfigure{\includegraphics[width=4.06cm]{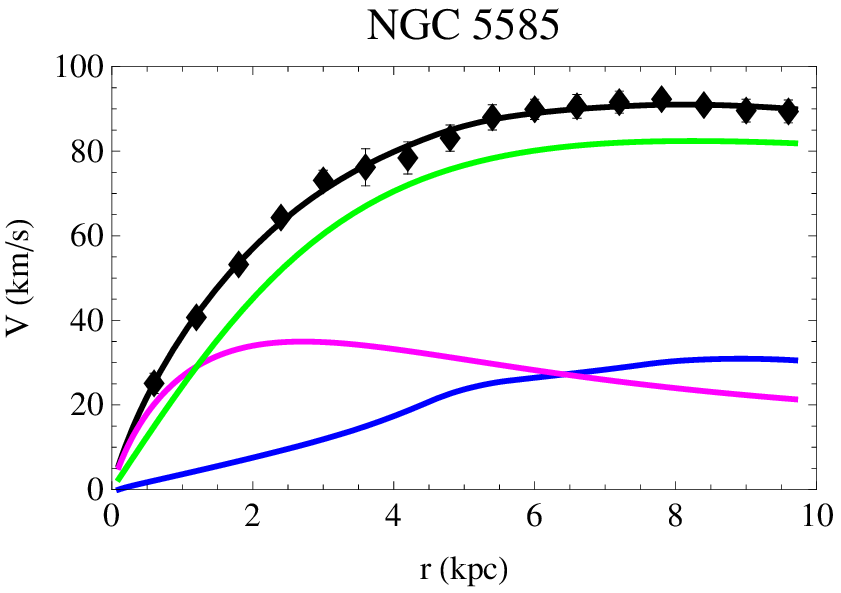}}\goodgap
\subfigure{\includegraphics[width=4.06cm]{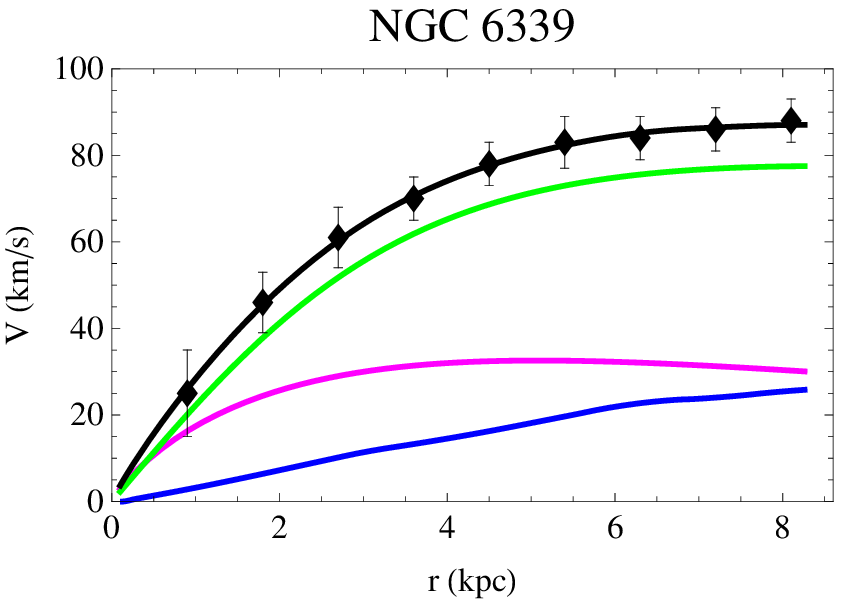}}\goodgap\\
\subfigure{\includegraphics[width=4.06cm]{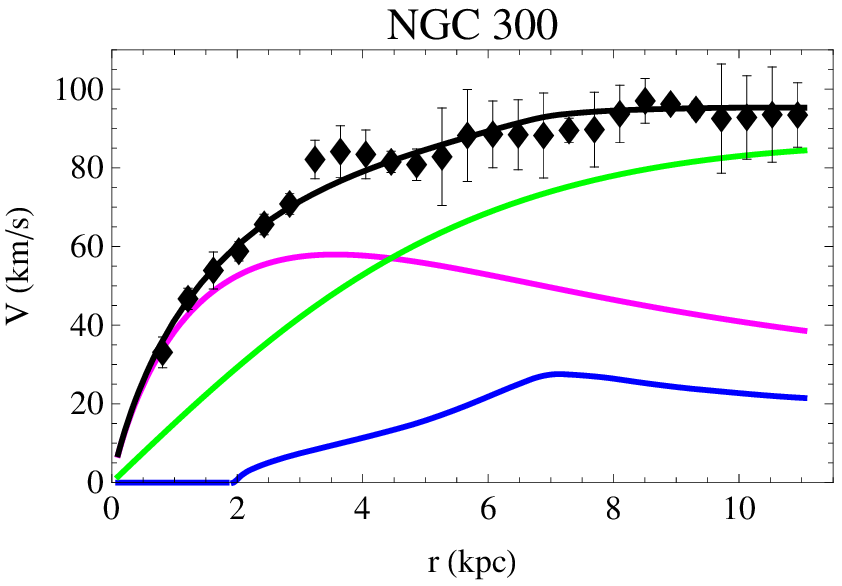}}\goodgap
\subfigure{\includegraphics[width=4.06cm]{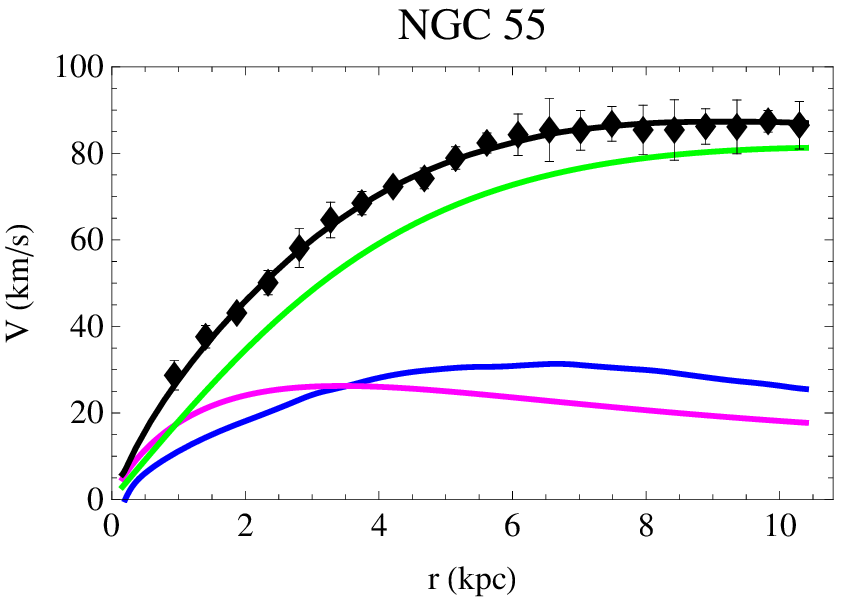}}\goodgap
\subfigure{\includegraphics[width=4.06cm]{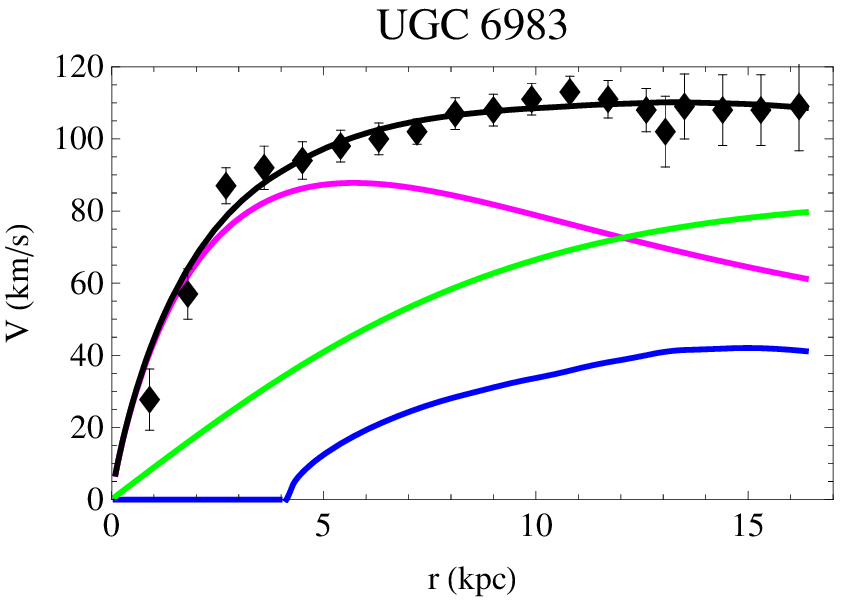}}\goodgap
\subfigure{\includegraphics[width=4.06cm]{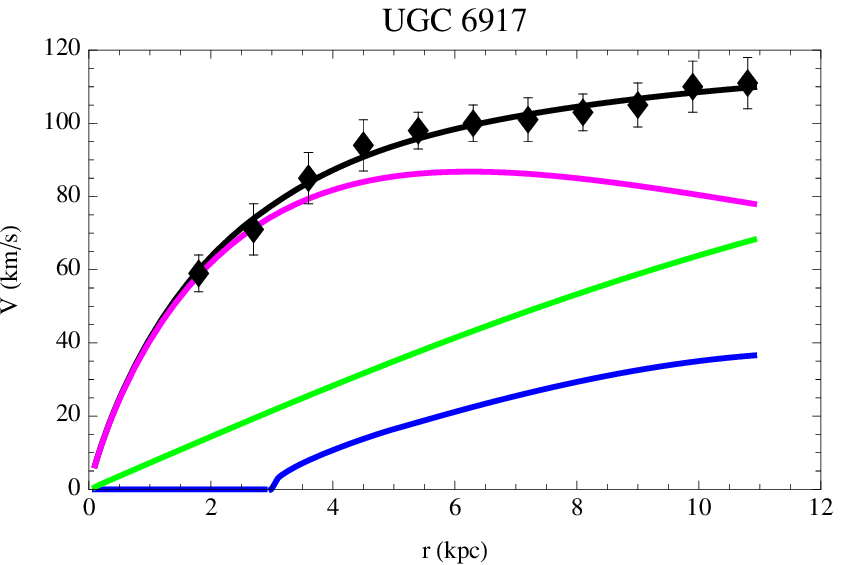}}\goodgap\\
\subfigure{\includegraphics[width=4.06cm]{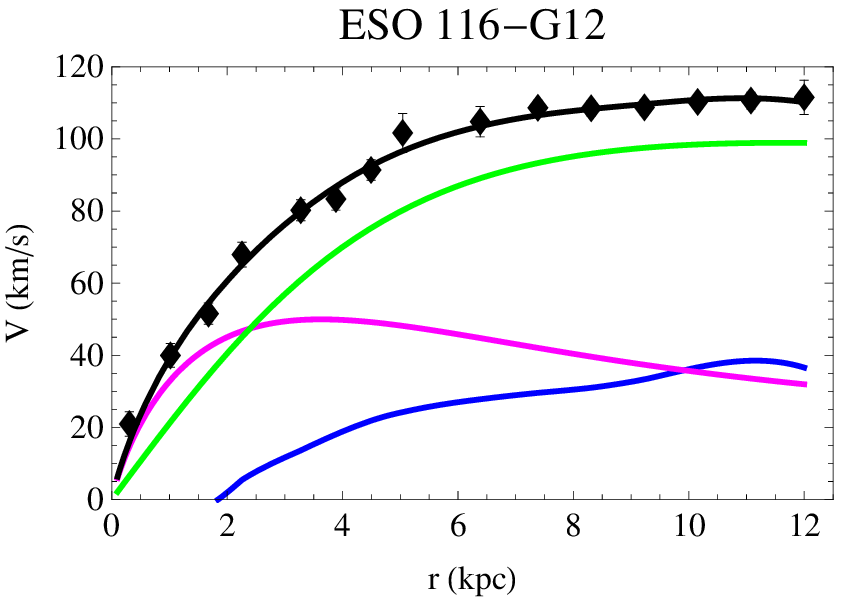}}\goodgap
\subfigure{\includegraphics[width=4.06cm]{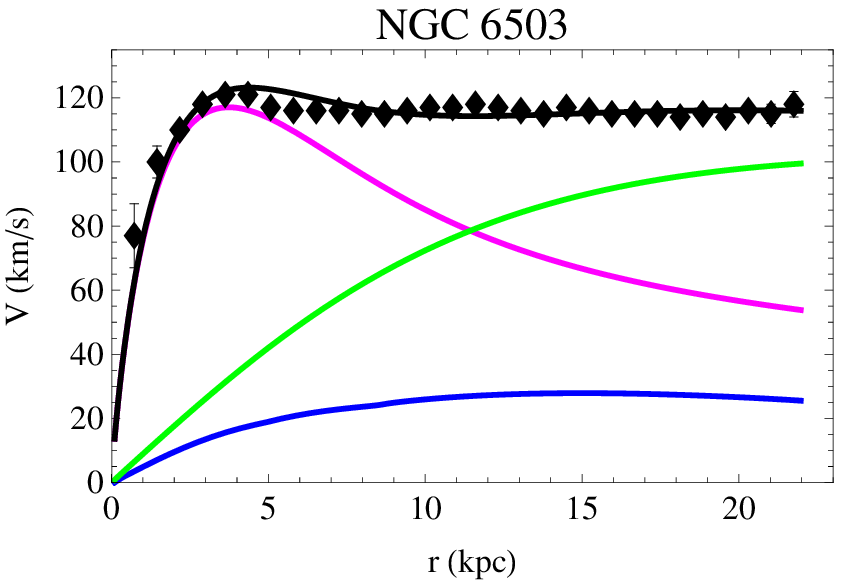}}\goodgap
\subfigure{\includegraphics[width=4.06cm]{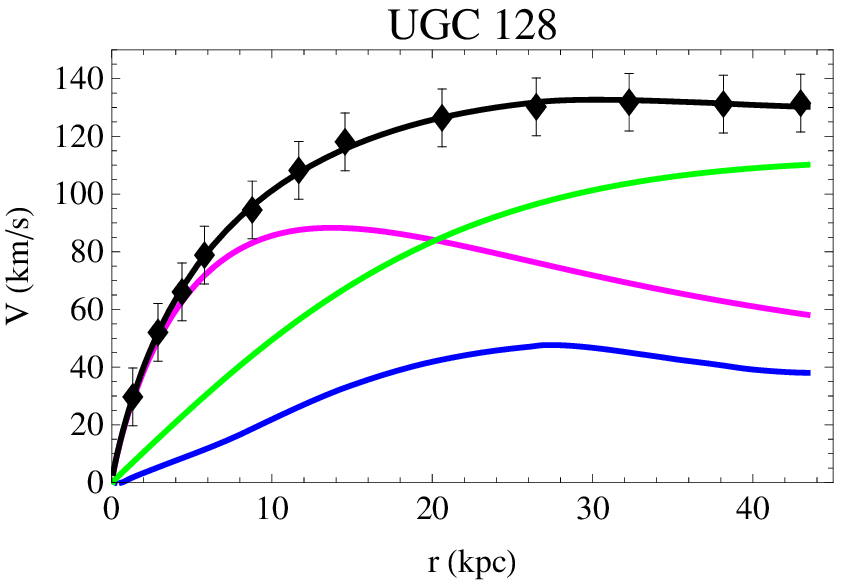}}\goodgap
\subfigure{\includegraphics[width=4.06cm]{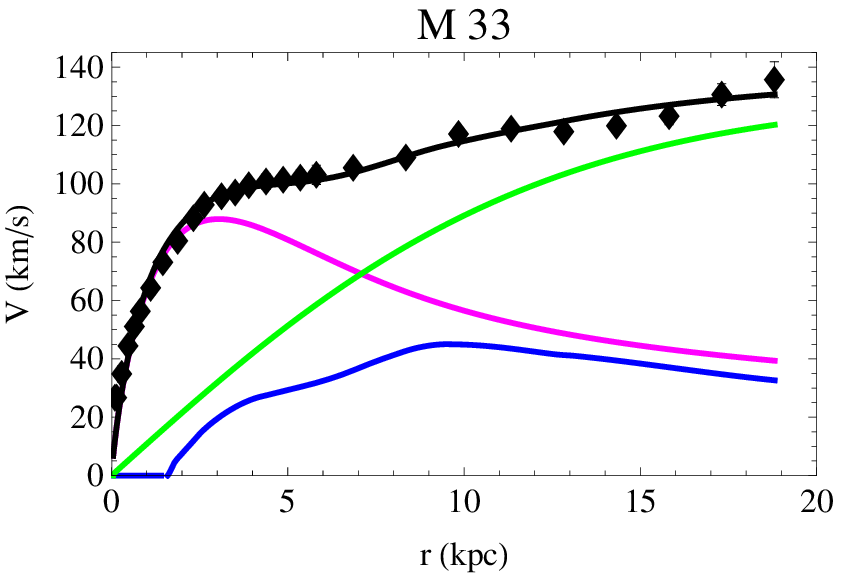}}\goodgap\\
\subfigure{\includegraphics[width=4.06cm]{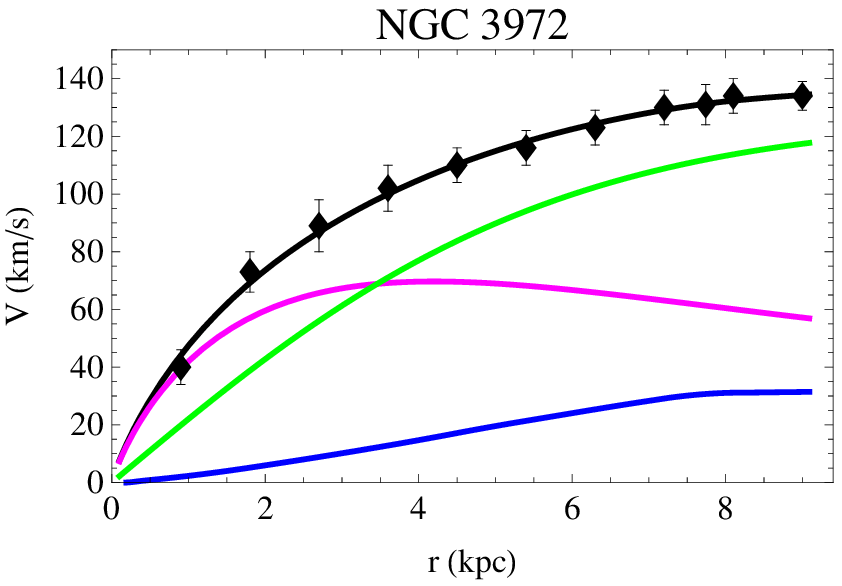}}\goodgap
\subfigure{\includegraphics[width=4.06cm]{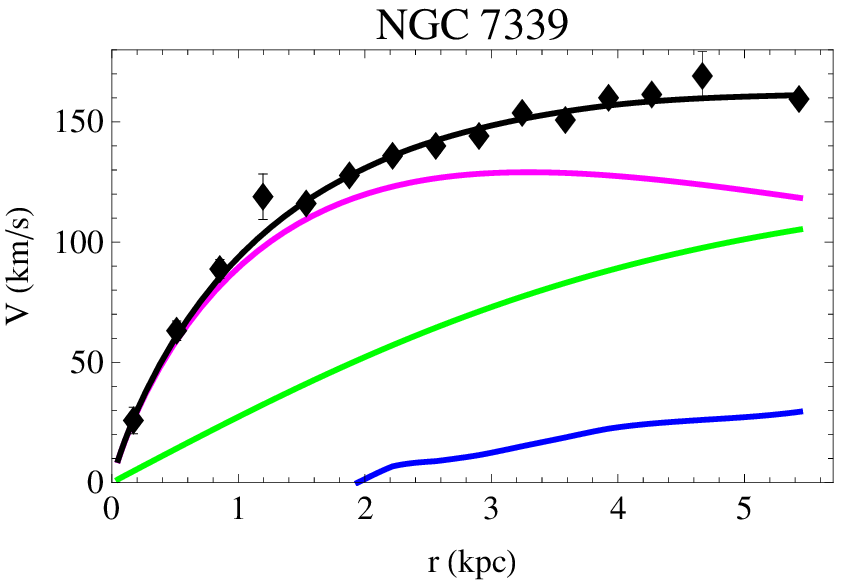}}\goodgap
\subfigure{\includegraphics[width=4.06cm]{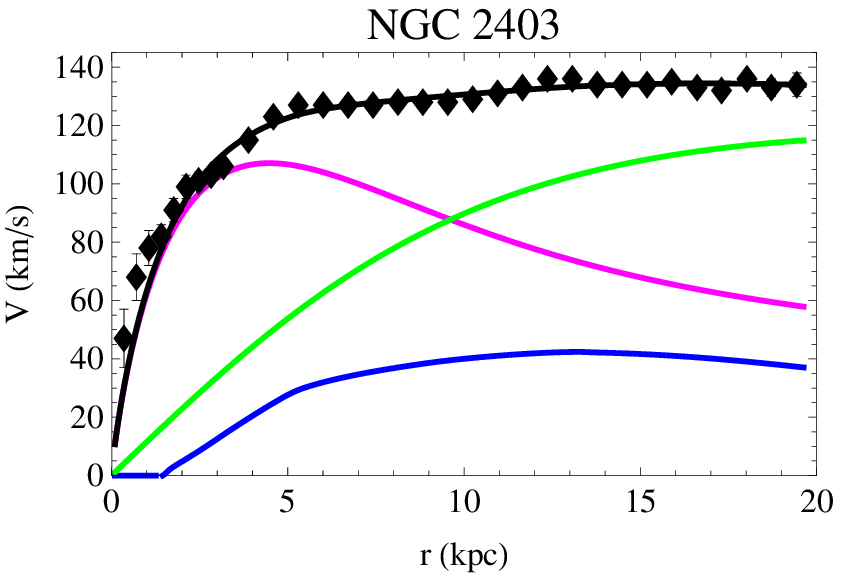}}\goodgap
\subfigure{\includegraphics[width=4.06cm]{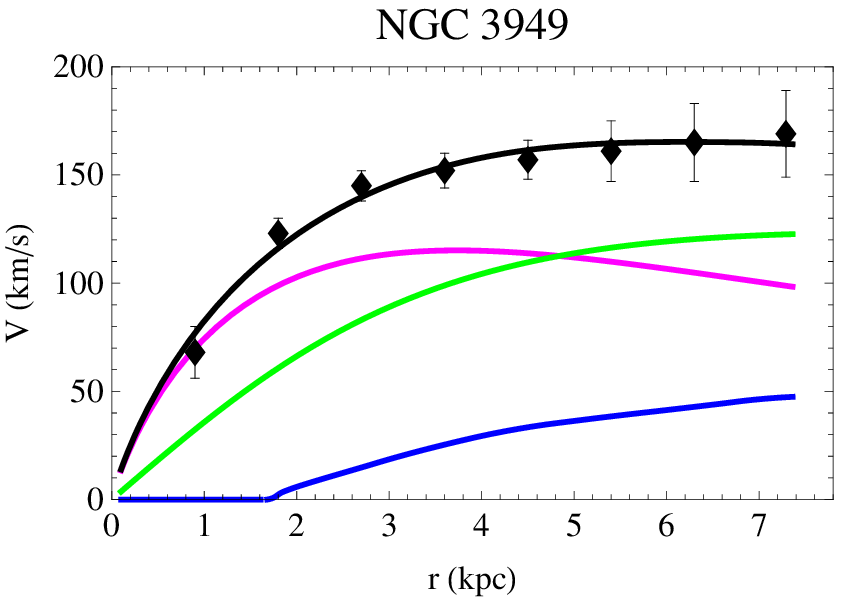}}\goodgap\\
\subfigure{\includegraphics[width=4.06cm]{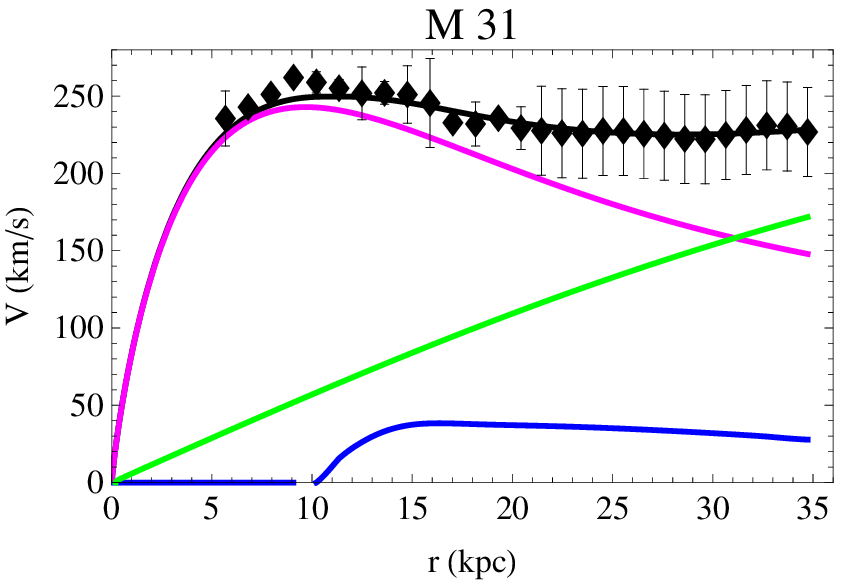}}\goodgap
\subfigure{\includegraphics[width=4.06cm]{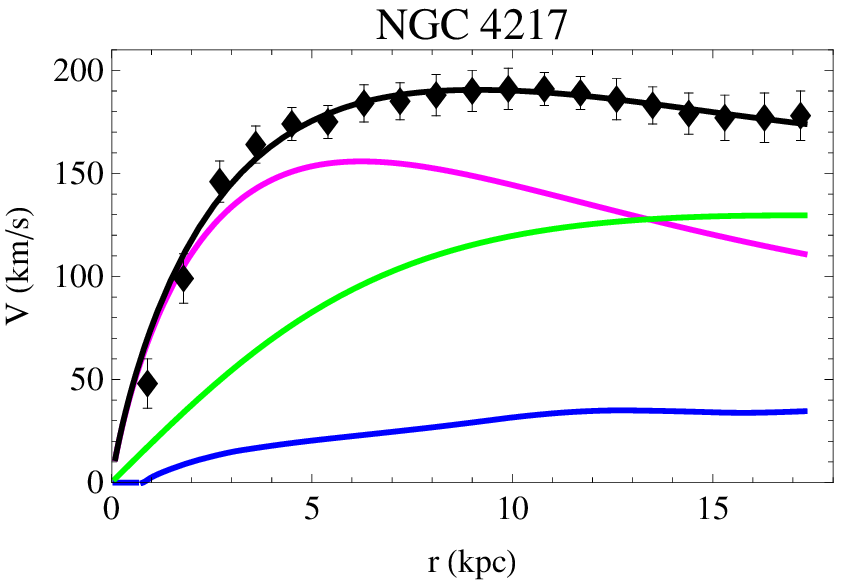}}\goodgap
\subfigure{\includegraphics[width=4.06cm]{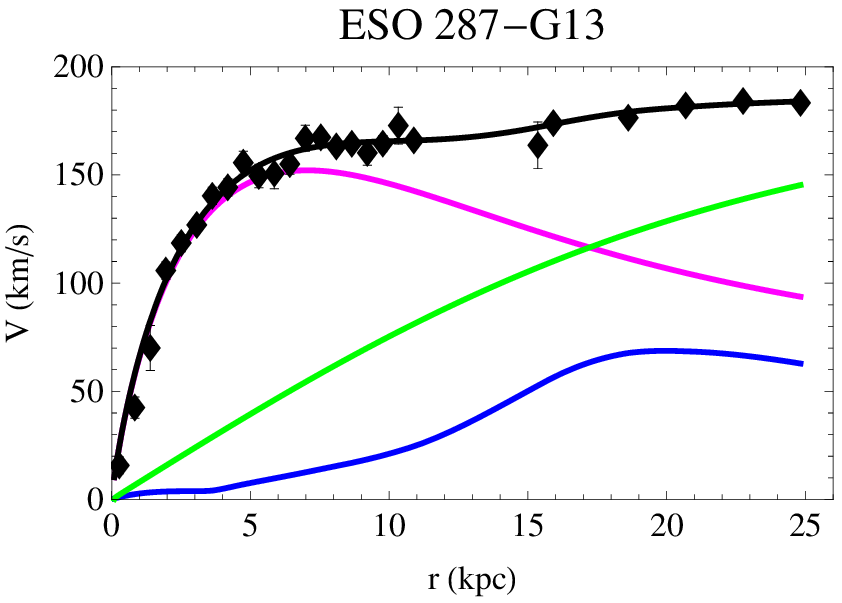}}\goodgap
\subfigure{\includegraphics[width=4.06cm]{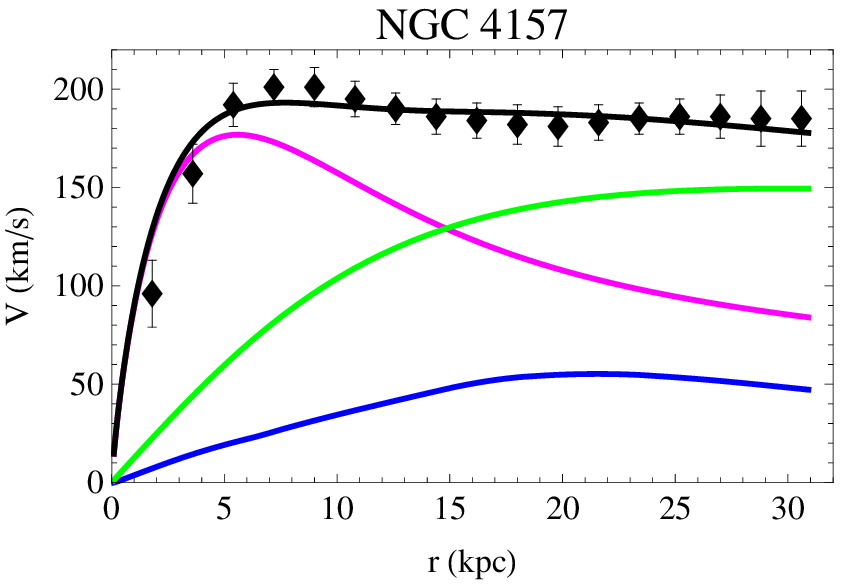}}\goodgap\\
\subfigure{\includegraphics[width=4.06cm]{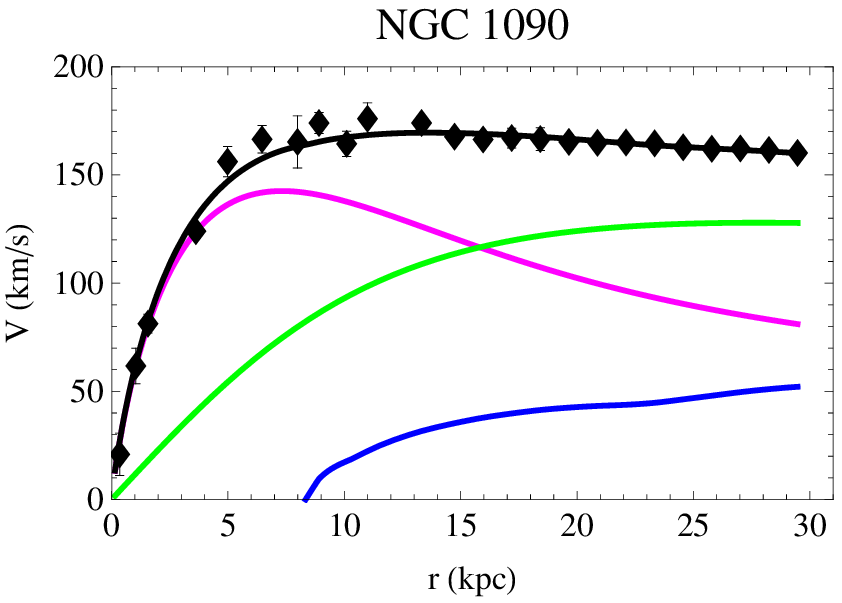}}\goodgap
\subfigure{\includegraphics[width=4.06cm]{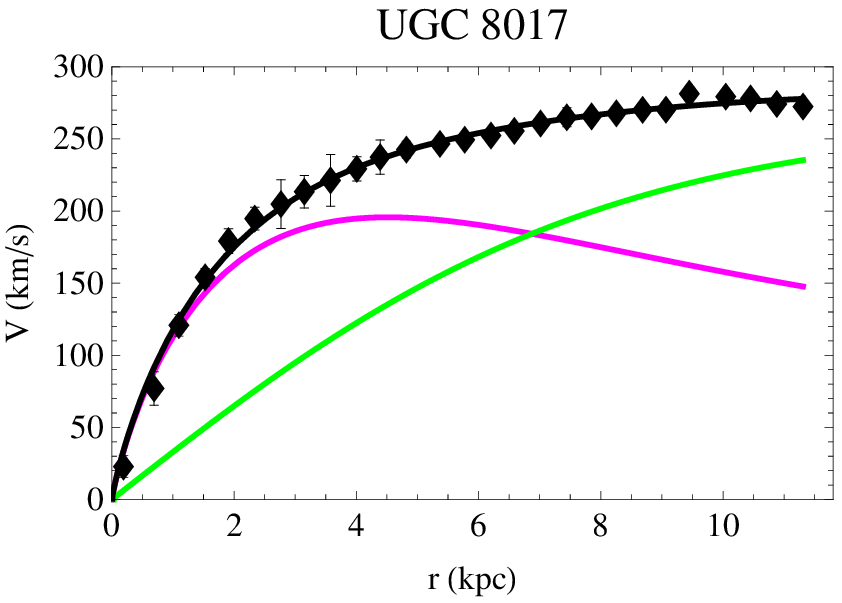}}\goodgap
\subfigure{\includegraphics[width=4.06cm]{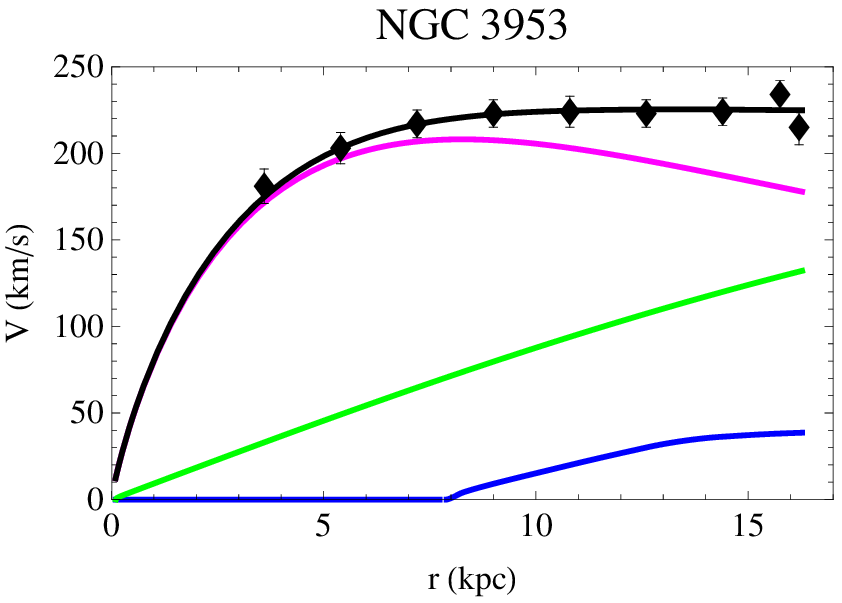}}\goodgap
\subfigure{\includegraphics[width=4.06cm]{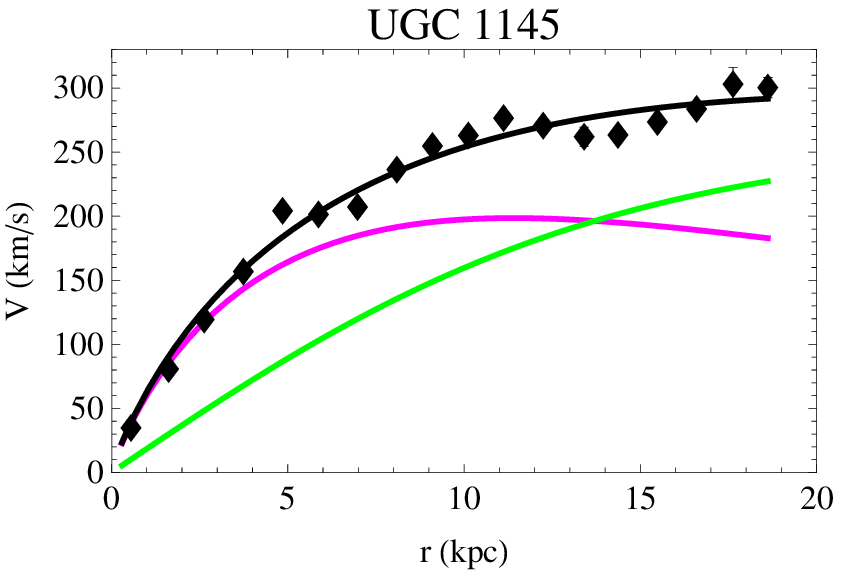}}\goodgap
\caption{Best-fitting curves superimposed to the data from selected objects from Frigerio Martins $\&$ Salucci (2007).
The black line represents the total circular velocity $V_{tot}$.
The blue and magenta lines are the contributions from the gaseous and stellar disks, while the green is the DM halo contribution.
See Table 1 for more details.}
\label{fig:bf}
\end{figure*}

\section{Conclusions}

In this article we have examined the spiral rotation curves and the associated mass models assuming that the  
$q$-statistics (and the associated kinetic phase space description)  provides the correct theoretical treatment responsible for the dark matter halo distribution.  
It was shown that a set of polytropic,  non-Gaussian, Lane-Emden Spheres with  the central value $q = 0.85$  yielded a successfully fitting for all the observed rotation curves  (in our statistical analysis $q = 0.85 \pm 0.35 (1\sigma)$).  It is worth notice that the allowed range of the $q$-index includes galactic systems with finite and infinite total mass. It has been proven that the mass is finite only if $q \leq 5/7$
\cite{lima05}.

In this concern, we would also like to call attention for the interest to consider the class of nonextensive truncated models for dark matter halos. In this case, only objects with energy per unit mass $\epsilon \leq 0$ are present in the distribution.
The density profile of these truncated models  can  readily be derived by numerical integration involving  the Poisson equation and the density as a function of the potential
\cite{lima05}. It is interesting to compare the predictions based on both approaches, but, naturally,  such a problem  lies beyond the scope of the present paper. A detailed analysis will be presented in a forthcoming communication. 

Finally, it should be stressed that the development of this field in the last two decades is strongly suggesting that time is ripe to meet a dark halo profile of galaxies inspired by fundamental principles.

\vspace{0.3cm}
\textit{Acknowledgments.} This work was partially supported by CNPq (JASL) and FAPESP (CFM).

\begin{table}
\caption{Best-fitting results: parameters and associated errors.
Galaxies are ordered from top to bottom with increasing luminosity.}
\begin{center}
\begin{tabular}{l|r|r|r|c|c}\hline\hline
\emph{Galaxy}&\emph{$M_D (10^{9} M_{\odot})$}&\emph{$r_o (kpc)$}&\emph{$\rho_q (10^{-24} g/cm^3)$}\\ \hline
DDO 47&0.06$^{+0.07}_{-0.06}$ &2.2$^{+4.5}_{-0.7}$  &1.2$^{+0.5}_{-0.4}$  \\\hline
IC 2574&0.4$^{+0.5}_{-0.4}$  &3.6$^{>4.9}_{-1.6}$&0.4$^{+0.3}_{-0.1}$  \\\hline
NGC 5585&0.9$^{+0.6}_{-0.7}$  &1.4$^{+0.3}_{-0.3}$&2.4 $^{+1.3}_{-0.8}$ \\\hline
UGC 6339&1.5$^{+6.2}_{-1.5}$  &1.5$^{+7.4}_{-0.2}$&2.0 $^{+1.6}_{-1.6}$ \\\hline
NGC 300&3.3$^{+1.5}_{-2.9}$ &2.5$^{>9.5}_{-1.3}$&0.9$^{+3.0}_{-0.6}$  \\\hline
NGC 55&0.7$^{+0.8}_{-0.7}$  &2.0$^{+0.8}_{-0.5}$&1.3$^{+0.8}_{-0.5}$  \\\hline
UGC 6983&12.2$^{+4.2}_{-12.2}$&4.1$^{>12.9}_{-2.9}$&0.3$^{+4.9}_{-0.2}$\\\hline
UGC 6917&13.0 $^{+3.7}_{-13.0}$ &6.7$^{>5.3}_{-5.3}$&0.2$^{+5.6}_{-0.1}$  \\\hline
ESO 116-G12&2.5$^{+1.5}_{-1.8}$  &2.0$^{+0.8}_{-0.5}$&1.7$^{+1.5}_{-0.8}$  \\\hline
NGC 6503&14.2$^{+1.0}_{-1.0}$  &5.0$^{+1.3}_{-0.9}$&0.3$^{+0.1}_{-0.1}$ \\\hline
UGC 128&29.7$^{+17.6}_{-29.7}$  &9.3$^{>40.7}_{-5.8}$&0.1$^{+0.7}_{-0.1}$ \\\hline
M 33&6.5$^{+0.5}_{-0.5}$  &5.2$^{+1.5}_{-1.0}$&0.4$^{+0.1}_{-0.1}$  \\\hline
NGC 3972&5.6$^{+4.8}_{-4.6}$  &2.5 $^{>7.5}_{-0.9}$&1.9$^{+4.0}_{-1.3}$ \\\hline
NGC 7339&15.2$^{+3.0}_{-4.1}$  &1.9$^{>48.1}_{-0.9}$&2.9 $^{+7.2}_{-1.9}$ \\\hline
NGC 2403&14.2$^{+1.7}_{-1.9}$  &4.5$^{+1.7}_{-1.1}$&0.5 $^{+0.3}_{-0.2}$\\\hline
NGC 3949&13.7$^{+1.6}_{-13.7}$  &1.5$^{+2.5}_{-0.6}$&5.1$^{+4.9}_{-3.3}$  \\\hline
M 31&158.0$^{+22.0}_{-4.0}$&20.5$^{>14.5}_{-9.0}$&0.1 $^{+0.1}_{-0.1}$ \\\hline
NGC 4217&41.9$^{+16.5}_{-41.9}$  &3.0$^{>15.1}_{-1.6}$&1.4 $^{+3.7}_{-1.3}$\\\hline
ESO 287-G13&45.2$^{+4.5}_{-5.2}$  &9.7$^{>20.3}_{-3.7}$&0.2$^{+0.2}_{-0.1}$  \\\hline
NGC 4157&48.4 $^{+14.7}_{-29.3}$ &5.2$^{+8.9}_{-2.9}$&0.6$^{+13.0}_{-0.4}$  \\\hline
NGC 1090&41.2$^{+10.0}_{-12.6}$  &4.9$^{+2.5}_{-1.6}$&0.5 $^{+0.8}_{-0.3}$ \\\hline
UGC 8017&47.8$^{+3.1}_{-3.1}$&3.4 $^{+0.6}_{-0.4}$&4.1$^{+1.0}_{-0.8}$  \\\hline
NGC 3953&99.0$^{+23.9}_{-80.8}$&10.7$^{>7.3}_{-8.6}$&0.3$^{+6.9}_{-0.2}$  \\\hline
UGC 11455&124.0$^{+26.3}_{-35.9}$&6.0$^{+10.4}_{-2.0}$&1.3$^{+1.4}_{-0.7}$\\\hline
\end{tabular}
\end{center}
\end{table}

\vspace{0.2cm}
\begin{figure}
\centering
\vspace{-1.7 cm}
\includegraphics[width=6.5cm]{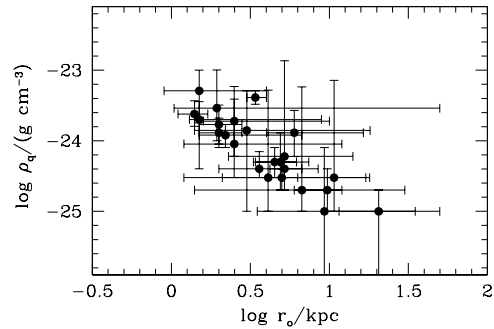}
\caption{Rotation curves and nonextensivity.
Parameters of $q$-statistics DM halos obtained from the analysis of rotation curves. }
\label{fig:bf}
\end{figure}

\begin{figure}
\centering
\includegraphics[width=6.5cm]{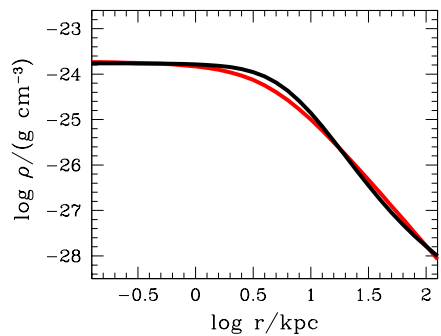}
\caption{Comparing DM density profiles.
The black curve represents the predicted halo density distribution based on the nonextensive approach
for the galaxy ESO 116-G12, while the red one is obtained from the Universal Rotation Curve.
Both plots were obtained for the same virial mass (see Figure 1 and Table 1).}
\end{figure}


\begin{thebibliography}{99}

\bibitem[\protect\citename{Abelev \emph{et al.}} 2013]{abelev13}
Abelev, B. \emph{et al.} (ALICE Collaboration), 2013, PRL 111, 222301

\bibitem[\protect\citename{Athanassoula, Bosma \& Papaioannou} 1987]{athanassoula87}
Athanassoula, E., Bosma, A., Papaioannou, S., 1987, A\&A 179, 23

\bibitem[\protect\citename{Betzler \& Borges} 2012]{betzler12}
Betzler, A. S., Borges, E. P., 2012, A\&A, 539, A158

\bibitem[\protect\citename{de Blok, McGaugh \& Rubin} 2001]{deBlok01}
de Blok, W. J. G., McGaugh, S. S., Rubin, V. C., 2001, AJ 122, 2396

\bibitem[\protect\citename{de Blok \& Bosma} 2002]{deBlok02}
de Blok, W. J. G., Bosma, A., 2002, A\&A 385, 816

\bibitem[\protect\citename{de Blok} 2010]{deBlok10}
de Blok W. J. G., 2010, AdAst, 2010, 789293

\bibitem[\protect\citename{Bosma \& van der Kruit} 1979]{bosma79}
Bosma, A., van der Kruit, P. C., 1979, A\&A 79, 281

\bibitem[\protect\citename{Bosma} 1981]{bosma81}
Bosma, A., 1981, AJ 86, 1825

\bibitem[\protect\citename{Burlaga \emph{et al.}} 2006]{burlaga06}
Burlaga, L. F., Vinas, A.F., Ness, N.F., Acuna, M. H., 2006,  Astrophys. J. Lett. 644, L83 

\bibitem[\protect\citename{Burkert} 1995]{burkert95}
Burkert A., 1995, ApJ, 447, L25 


\bibitem[\protect\citename{Cardone, Leubner \& Del Popolo} 2011]{cardone11}
Cardone, V. F., Leubner, M.P., Del Popolo, A., 2011, MNRAS 414, 2265

\bibitem[\protect\citename{Donato \emph{et al.}} 2009]{donato09}
Donato, F., Gentile, G., Salucci, P., Frigerio Martins, C., Wilkinson, M. I., Gilmore, G., Grebel, E. K., Koch, A., Wyse, R., 2009, MNRAS, 397, 1169 

\bibitem[\protect\citename{Du Jiulin} 2004]{duJiulin04}
Du Jiulin, 2004,  Europhys. Lett. 67, 893

\bibitem[\protect\citename{Freeman} 1970]{freeman}
Freeman, K. C., 1970, ApJ 160, 811

\bibitem[\protect\citename{Frigerio Martins \& Salucci} 2007]{frigerio}
Frigerio Martins, C., Salucci, P., 2007, PRL 98, 151301

\bibitem[\protect\citename{Gentile \emph{et al.}} 2004]{gentile04}
Gentile, G., Salucci, P., Klein, U., Vergani, D., Kalberla, P., 2004, MNRAS 351

\bibitem[\protect\citename{Gentile \emph{et al.}} 2005]{gentile05}
Gentile, G., Burkert, A., Salucci, P., Klein, U., Walter, F., 2005, ApJ 634, L145

\bibitem[\protect\citename{Gentile \emph{et al.}} 2005]{47}
Gentile, G. \emph{et al.}, 2005, ApJ 234, L145

\bibitem[\protect\citename{Gentile \emph{et al.}} 2007]{gentile07a}
Gentile, G., Salucci, P., Klein, U., Granato, G. L., 2007, MNRAS 375, 199

\bibitem[\protect\citename{Gentile, Tonini \& Salucci} 2007]{gentile07b}
Gentile, G., Tonini, C., Salucci, P., 2007,  A\&A 467, 925

\bibitem[\protect\citename{Hansen \emph{et al.}} 2005]{hansen05} Hansen, S. H., Egli, D., Hollenstein, 
L. \& Salzmann, C., 2005, New Astronomy 10, 379

\bibitem[\protect\citename{Hansen} 2009]{hansen09}
Hansen, S. H. \& Moore, B., 2006, New Astronomy 11,  333 

\bibitem[\protect\citename{Kaniadakis} 2005]{kaniadakis05}
Kaniadakis, G., Lissia, M., Scarfone, A.M., 2005,  Phys. Rev. E 71, 046128

\bibitem[\protect\citename{Kronenberger, Leubner \& van Kampen} 2006]{kronenberger06}
Kronenberger T., Leubner M. P., van Kampen E., 2006, A\&A, 453, 21

\bibitem[\protect\citename{Leubner} 2005]{leubner05}
Leubner, M. P., 2005, ApJ, 632, L1

\bibitem[\protect\citename{Lima \emph{et al.}} 2000]{lima00}
Lima, J. A. S., Silva, R., Santos, J., 2000, Phys. Rev. E 61, 3260

\bibitem[\protect\citename{Lima, Plastino \& Silva} 2001]{lima01}
Lima, J. A. S., Plastino, A. R., Silva, R., 2001, PRL 86, 2938 

\bibitem[\protect\citename{Lima, Silva \& Santos} 2002]{lima02}
Lima, J. A. S., Silva, R. \& Santos, J., 2002, A\&A 396, 309

\bibitem[\protect\citename{Lima \& de Souza} 2005]{lima05}
Lima, J. A. S., de Souza, R. E., 2005, Physica A 350, 303

\bibitem[\protect\citename{Marchesini \emph{et al.}} 2002]{marchesini02}
Marchesini, D., D'Onghia, E., Chincarini, G., Firmani, C., Conconi, P., Molinari, E., Zacchei, A., 2002, ApJ 575, 801

\bibitem[\protect\citename{Moore \emph{et al.}} 1999]{moore99}
Moore, B., Quinn, T., Governato, F., Stadel, J., Lake, G., 1999, MNRAS 310, 1147

\bibitem[\protect\citename{Kuzio de Naray \& Spekkens} 2011]{naray11}
Kuzio de Naray, R., Spekkens, K., 2011, ApJ, 741, L29 

\bibitem[\protect\citename{Navarro, Frenk \& White} 1996]{nfw96}
Navarro, J. F., Frenk, C.S., White, S.D.M., 1996, ApJ 462, 563

\bibitem[\protect\citename{Navarro \emph{et al.}} 2004]{navarro04}
Navarro, J. F., Hayashi, E., Power, C., Jenkins, A. R., Frenk, C. S., White, S. D. M., Springel, V., Stadel, J., Quinn, T. R., 2004, MNRAS 349, 1039

\bibitem[\protect\citename{Neto \emph{et al.}} 2007]{neto07}
Neto, A. F., Gao, L., Bett, P., Cole, S., Navarro, J. F., Frenk, C. S., White, S.D.M., Springel, V., Jenkins, A., 2007, 
MNRAS 381, 1450

\bibitem[\protect\citename{Palunas \& Williams} 2000]{palunas00}
Palunas, P., Williams, T. B., 2000, AJ 120, 2884

\bibitem[\protect\citename{Persic \& Salucci} 1988]{persic88}
Persic, M., Salucci, P., 1988, MNRAS 234, 131

\bibitem[\protect\citename{Persic \& Salucci} 1991]{persic91}
Persic, M., Salucci, P., 1991, ApJ 368, 60

\bibitem[\protect\citename{Persic, Salucci \& Stel} 1996]{persic96}
Persic, M., Salucci, S., Stel, F., 1996, MNRAS 281, 27

\bibitem[\protect\citename{Rubin, Thonnard \& Ford} 1980]{rubin80}
Rubin, V. C., Thonnard, N., Ford, W. K. Jr., 1980, ApJ 238, 471

\bibitem[\protect\citename{Salucci \& Persic} 1999]{salucci99}
Salucci, P., Persic, M., 1999, A\&A, 351, 442

\bibitem[\protect\citename{Salucci \& Burkert} 2000]{salucci00}
Salucci, P., Burkert, A., ApJ, 2000, 537, L9

\bibitem[\protect\citename{Salucci \emph{et al.}} 2007]{salucci07}
Salucci, P., Lapi, A., Tonini, C., Gentile, G. , Yegorova, I.A., Klein, U., 2007, MNRAS 378, 41

\bibitem[\protect\citename{Salucci, Yegorova \& Drory} 2008]{salucci08}
Salucci P., Yegorova I. A., Drory, N., 2008, MNRAS, 388, 159 

\bibitem[\protect\citename{Salucci \emph{et al.}} 2012]{salucci12} Salucci P., Wilkinson M.I ., Walker M. G., 
Gilmore G. F., Grebel E. K., Koch A., Frigerio Martins, C., Wyse R. F. G., 
2012, MNRAS, 420, 2034 

\bibitem[\protect\citename{Silva, Alcaniz, lima} 2004]{silva04} Silva R., Alcaniz J. S., Lima, J. A. S.,  2005, Physica A 356, 509

\bibitem[\protect\citename{Silva, Plastino \& Lima} 1998]{silva98} Silva, R., Plastino, A. R., Lima, J. A. S., 1998, Phys. Lett. A 249, 401

\bibitem[\protect\citename{Silva \& Lima} 2005]{silva05} Silva, R., Lima, J. A. S., 2005, Phys. Rev. E 72, 057101 

\bibitem[\protect\citename{Taruya \& Sakagami} 2002]{taruya02} Taruya, A., Sakagami, M., 2002, Physica A, 307, 185

\bibitem[\protect\citename{Tsallis} 1988]{T88} Tsallis, C., 1988, J. Stat. Phys. 52, 479

\bibitem[\protect\citename{Tsallis} 2014]{tsallis14}
Tsallis, C., 2014, EPJ WEB Coferences 71, 00132

\bibitem[\protect\citename{Wong \& Wilk} 2013]{wong13}
Wong, C.-Y., Wilk, G., 2013, Phys. Rev. D 87, 114007


\end{thebibliography}
\end{document}